\documentclass[preprint,prb]{revtex4}
\usepackage{stmaryrd}
\usepackage{amsmath}
\usepackage{amssymb}
\usepackage{graphicx}
\usepackage{dcolumn}
\usepackage{bm}

\begin{document}

\begin{titlepage}

\title{Dirac Fermions in Blue-Phosphorus}
\author{Yuanchang Li,$^{1}$\footnote{liyc@nanoctr.cn} and Xiaobin Chen$^{2,3}$}
\address{$^1$National Center for Nanoscience and Technology, Beijing 100190, People¡¯s Republic of China \\
$^2$Department of Physics and State Key Laboratory of Low-Dimensional Quantum Physics, Tsinghua University, Beijing 100084, People's Republic of China \\
$^3$Collaborative Innovation Center of Quantum Matter, Tsinghua University, Beijing 100084, China}

\date{\today}

\begin{abstract}
We propose that Dirac cones can be engineered in phosphorene with fourfold-coordinated phosphorus atom. The key is to separate in energy the in-plane ($s$, $p_x$ and $p_y$) and out-of-plane ($p_z$) oribtals through the $sp^2$ configuration, yielding respective $\sigma$- and $\pi$-character Dirac cones, and then quench the latter. As a proof-of-principle study, we realize $\sigma$-character Dirac cone in hydrogenated/fluorinated phosphorene with the honeycomb lattice. The obtained Dirac cones are at $K$-points, slightly anisotropic, with Fermi velocities of 0.91/1.23 times that of graphene along $\Gamma$K/KM direction, and maintain a good linearity up to $\sim$2 eV for holes. One substantive advantage of $\sigma$-character Dirac cone is its convenience to tune the Dirac gap via in-plane strain. Our findings pave a new way for development of high performance electronic devices based on Dirac materials.
\end{abstract}

\maketitle

\draft

\vspace{2mm}

\end{titlepage}

\section{Introduction}
During the last decade, graphene has attracted tremendous attention and research interests owing to its possible applications in carbon-based nanoelectronics.\cite{Novoselov,Avouris,Loh} Graphene has the $\pi$-character Dirac cone: $\pi$ and $\pi^\ast$ bands, contributed by the carbon out-of-plane $p_z$ orbitals, intersect with each other at the $K$ and $K^\prime$ points, respectively. The band structure is directly related to its honeycomb lattice symmetry, but unfortunately, the unique symmetry is usually disturbed or destroyed when graphene is coupled to the external world in device fabrication, e.g., bonding with the substrate.\cite{usprl} On the other hand, opening a tunable bandgap in graphene is still a big challenge, which is essential for controlling the conductivity by electronic means in the semiconductor industry. Owing to the out-of-plane nature in graphene, the homogeneous strain behaves like the effective electronic scalar potential and only particularly designed strain distribution can gap the Dirac cone.\cite{Guinea,Pereira,Low} These raise an interesting question: whether there exists the $\sigma$-character Dirac cone spectrum, originated from the in-plane $p_x$ and $p_y$ orbitals. Such an in-plane nature will not only enrich the Dirac physics in fundamental science but also may open a new possibility for the gap engineering in Dirac materials via the in-plane strain for their technological applications.

Phosphorene is another stable two-dimensional elemental matter besides graphene but has an inherent bandgap.\cite{Reich,LiuH,LiL,Castellanos-Gomez,Rudenko,Fei,Koenig,Xia,blueZhu,Yao,Peng,Zeng,Qiao} Recently, monolayer phosphorene is exfoliated\cite{LiuH}, which also possesses a hexagon skeleton like graphene. Phosphorus has several stable allotropes with different bandgaps, among which black phosphorus is the most stable. Unlike graphene, phosphorene exfoliated from black phosphorus takes a puckered non-planar structure although it is composed of the basic hexagon as shown in Fig. 1(a). Very recently, blue phosphorus, a new member of phosphorus family, has been proposed to share equal stability with the black one\cite{blueZhu}, from which the phosphorene has the buckled honeycomb lattice in analogy to silicene\cite{Silicene} (See Fig. 1(b)). For clarity, we will use blue phosphorene as a prototype to illustrate our concept of engineering Dirac cone spectrum in phosphorus and we expect that the same concept also works in any other phosphorus allotrope that possesses a $sp^2$ configuration.

In this paper, we perform a theoretical study of $\sigma$-character Dirac cone in a one-atom-thick phosphorene. We show that the (quasi-)planar honeycomb lattice can separate in energy the in-plane ($s$, $p_x$ and $p_y$) and out-of-plane ($p_z$) oribtals, resulting in the $\sigma$- and $\pi$-character Dirac cones, respectively. We then quench the $\pi$-character Dirac cone by out-of-plane saturation, leaving the $\sigma$-character Dirac cone behind. We take hydrogenated phosphorene as a model system to investigate the properties of the $\sigma$-character Dirac cone. Our obtained Dirac cones are at $K$-points, slightly anisotropic, with Fermi velocities of 0.91/1.23 times that of graphene along $\Gamma$K/KM direction, and maintain a good linearity up to $\sim$2 eV for holes. We demonstrate that a tunable bandgap can be engineered via the in-plane strain herein apart from the AB sublattice symmetry breaking. Finally, we discuss the possibility to incorporate the $\sigma$-character Dirac cone into the semiconductor industry through the fourfold-coordinated configuration, which also helps to overcome the instability of phosphorene when exposure to the air as well.

\begin{figure}[tbp]
\includegraphics[width=0.6\textwidth]{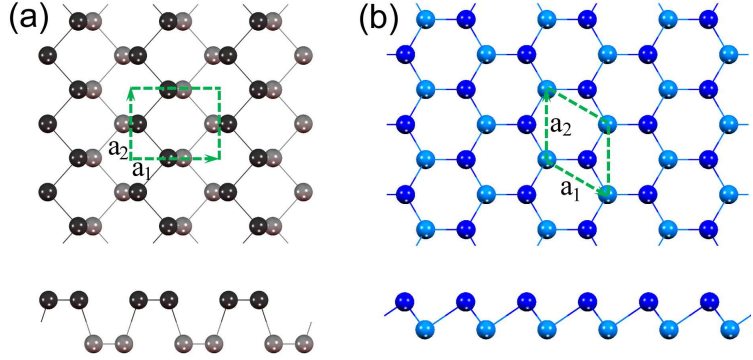}
\caption{\label{Figure 1} (a) Black and (b) blue monolayer phosphorene. Upper panel for the top view and the lower panel for the side view. Both phosphorenes have the hexagon skeleton like graphene but with different lattice symmetries (namely, orthorhombic for the former and hexagonal for the latter). Different shades of the atoms denote their different heights in the buckled structure. Note that monolayer black phosphorene has been exfoliated experimentally while the blue phosphorene was theoretically predicted to possess the same thermal stability as the black one.}
\end{figure}

\section{Methods and Model}
The calculations were performed using density-functional theory (DFT) with the projector augmented wave \cite{PAW} method and the Perdew-Burke-Ernzerhof \cite{PBE} exchange and correlation potential, as implemented in the Vienna \emph{ab initio} simulation package \cite{vasp}. An energy cutoff of 400 eV was employed. A gamma-centered grid of 36 $\times$ 36 $\times$ 1 was used to sample the Brillouin zone. The vacuum layer thickness was larger than 10 \AA.

\section{Results and Discussion}

\begin{figure}[tbp]
\includegraphics[width=0.7\textwidth]{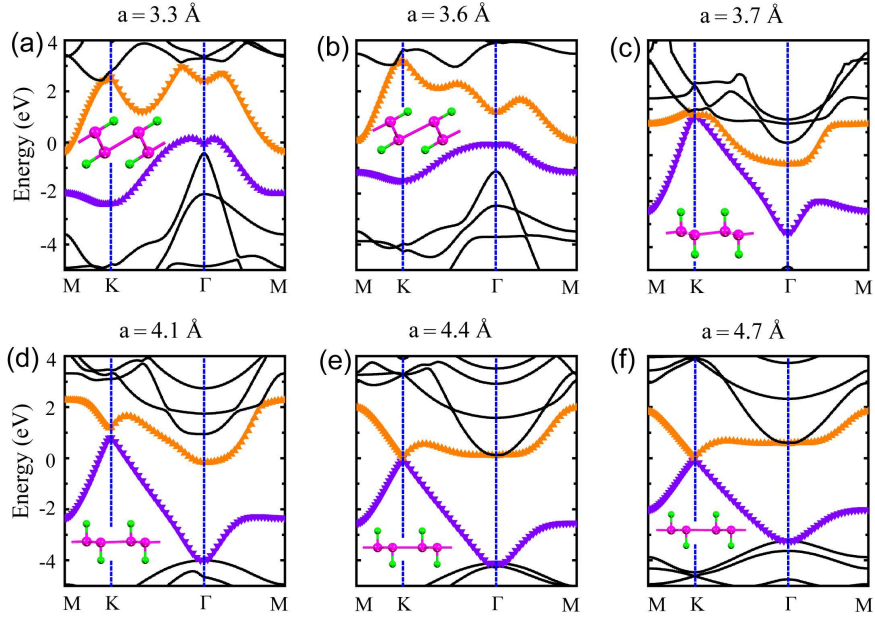}
\caption{\label{Figure 2} Band structures of double-side hydrogenated monolayer blue phosphorene at the typical lattice parameters $a$. The two bands ultimately evolving into the Dirac cone are highlighted in orange with up-triangle and violet with down-triangle as a guide to the eye. Fermi level is set to zero. Insets are side view of the configurations. Large magenta and small green balls denote the phosphorus and hydrogen atoms, respectively.}
\end{figure}

Figure 2 plots the geometric and electronic structures of double-side hydrogenated monolayer blue phosphorene with different lattice parameters ($a_1$=$a_2$=$a$). Test calculations show that there is no net spin-polarization in the present cases. We begin with $a$=3.3 \AA\ due to the lattice parameter of 3.326 \AA\ for monolayer blue phosphorene\cite{blueZhu}. Substantially different from the $\sim$2 eV bandgap of blue phosphorene, hydrogen-induced-metallization is found as reflected by Fig. 2(a). Structurally, phosphorene maintains a remarkably buckled hexagonal skeleton. Hydrogenation leads to phosphorus dimerization and there appear two kinds of P-P bond length, 2.20 and 2.59 \AA. With increasing $a$, the system metallicity is gradually suppressed while the P-P dimerization becomes stronger and stronger. Until $a$=3.6 \AA, the system shows a semimetal feature (See Fig. 2(b)).

When $a$=3.7 \AA, both the geometric and electronic structures change drastically as shown in Fig. 2(c). On the one hand, the hydrogenated phosphorene becomes nearly planar and dimerization is notably suppressed. On the other hand, interesting band crossing emerges at the $K$-point in sharp contrast to the original gapped behavior at all $k$ points. Further increasing $a$, the two highlighted bands continuously separate from the other bands and approach each other at the $K$-point while they are mutually exclusive, and ultimately merge into other bands at the $\Gamma$ point, as illustrated in Figs. 2(d)-2(f). In particular, when $a$=4.4 \AA, the system becomes a Dirac semimetal characterized by a $K$-point Dirac cone (note: the band crossing at the $\Gamma$ point is 0.09 eV above the Fermi level), and all P-P bond lengths become equivalent along with the disruption of the dimerization. Further increasing $a$, the $K$-point Dirac cone remains, while the crossing at the $\Gamma$ point shifts towards higher energy, e.g., it shifts to 0.56 eV when $a$=4.7 \AA. The trend holds until the hopping between near-neighboring phosphorus is weak enough, and then strong correlation effect dominates the properties. For example, the system with $a$=5.3 \AA\ does not have a Dirac cone any more.

Apparently, the lattice parameter increases as high as 33\% from $a$=3.3 \AA\ to $a$=4.4 \AA, however, the P-P bond length just elongates from 2.27 \AA\ of pristine blue phosphorene to 2.54 \AA\ in our hydrogenated phosphorene ($a$=4.4 \AA), corresponding to a stretch of 12\%. This is originated from the simultaneous release of the out-of-plane buckling. As we show below, the Dirac cone can be obtained even when the P-P bond is only elongated by 4.4\%. Hence, it is not the strain but the structural phase transition from non-planar to planar that plays a key role in the presence of exotic Dirac cone spectrum.

\begin{figure}[tbp]
\includegraphics[width=0.7\textwidth]{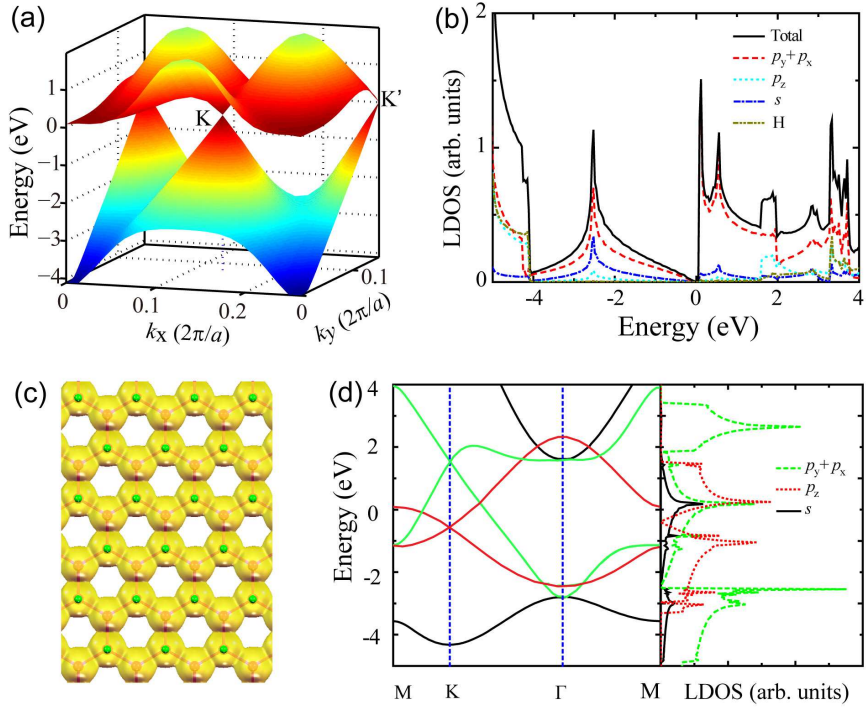}
\caption{\label{Figure 3} (a) Three-dimensional band structure around the Dirac point, (b) total and local density of states (LDOS), and (c) real-space charge distribution for the states in the energy range of [-1, 0] eV (isosurface = 0.0024 e/\AA$^3$) for double-side hydrogenated monolayer phosphorene with $a$=4.4 \AA, corresponding to Fig. 2(e). In (b) the contributions from phosphorus (P) $p_z$, $p_x$ + $p_y$, and $s$ orbitals, and hydrogen (H) are respectively shown. (d) Band structure and LDOS for the above system but with H removed. Red and green lines highlight the Dirac cones formed by P $p_z$ and $s$ + $p_x$ + $p_y$ orbitals, respectively. Fermi levels are at energy zero.}
\end{figure}

Following we take $a$=4.4 \AA\ case as an example to discuss the properties of the Dirac cone. Figure 3(a) shows the three-dimensional band structure around the Dirac point. The valley symmetry as in graphene is observed: two Dirac cones locate at the $K$ and $K^\prime$ points, respectively. Thus, a series of interesting valley-related phenomena might be realized in this system. Unlike in graphene, the energy spectrum is no longer electron-hole symmetric, and therefore neither the scattering mechanism nor transport properties would be identical for the electron and hole doping situations. The linear dispersion holds up to $\sim$2 eV for holes, while the massless electrons acquire mass rapidly away from the $K$-point. This implies that such a system is more promising to make unipolar ($p$-type) field effect devices rather than ambipolar ones with graphene. On the other hand, the linear dispersion region of electron can be widened by increasing $a$, e.g., it achieves 0.5 eV when $a$=4.7 \AA, the same as in graphene\cite{Wallace}. Moreover, the Dirac cone is slightly anisotropic, with the Fermi velocities of 0.91/1.23 times in graphene along $\Gamma$$K$/$KM$ direction. It is worthy emphasizing that the Fermi velocity depends weakly upon $a$ provided that the Dirac cone has formed.

Phosphorous atom has an intrinsically larger spin-orbit interaction than carbon atom. This would lead to many intriguing quantum phenomena in phosphorene, such as quantum spin Hall effect\cite{Kane}, besides the massless Dirac-fermion as in graphene. As shown below, the inversion symmetry is not necessary for the presence of Dirac cone any more and thus it is a natural candidate material for applications in valleytronics\cite{Rycerz}, and, with applied perpendicular electric field, quantum valley Hall effect as well\cite{Tahir}. Moreover, phosphorus is necessarily fourfold-coordinated herein, which provides us with a new freedom for property modulation and device fabrication.

Next we turn to study the physical origin of the unique Dirac cone. Figure 3(b) shows the local density of states (LDOS) of hydrogenated phosphorene with $a$=4.4 \AA, from which three hints are found. First, the Dirac cone dominantly arises from the phosphorus $p_x$, $p_y$ and $s$ orbitals rather than the $p_z$ orbitals as in graphene, clearly reflected by the LDOS within 2 eV below the Fermi energy. Generally, $p_x$, $p_y$ and $s$ orbitals form the in-plane $\sigma$ bonds, indicating an in-plane nature of the Dirac cone. The in-plane nature is further confirmed by the decomposed charge density for the states within 1 eV below the Fermi energy, as shown in Fig. 3(c). Second, strong hybridization between P $p_z$  and H states is observed in the energy region $\leq-4$ eV, indicating the passivation of $p_z$ states by H. Third, owing to the in-plane nature, the Dirac cone, not only the bandgap at the Dirac point but also the band shape, should be sensitive to the in-plane strain.

\begin{figure}[tbp]
\includegraphics[width=0.8\textwidth]{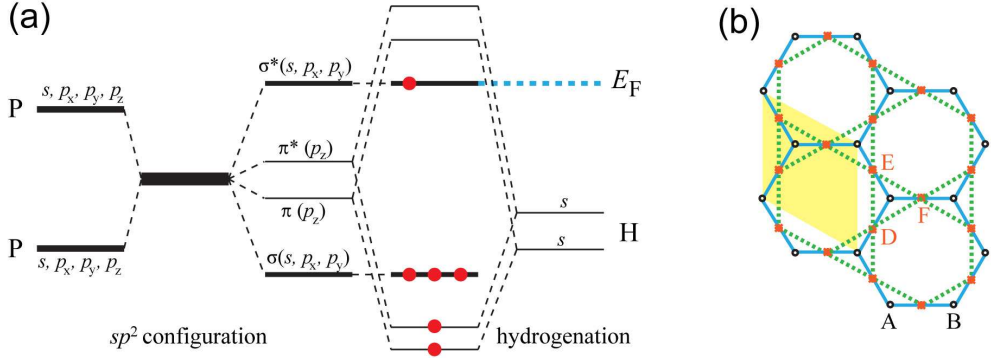}
\caption{\label{Figure 4} (a) On the honeycomb lattice, the $sp^2$ hybridization leads to bonding ($\sigma$ and $\pi$) and anti-bonding ($\sigma^\ast$ and $\pi^\ast$) orbitals similar to graphene. Hydrogenation adds two more electrons, which bond to the $p_z$ oribtals and quench the $\pi$-character Dirac cone, promoting one electron on three $\sigma^\ast$ levels. The line thickness represents the orbital degeneracy while the red dot represents an occupied electron. (b) Electron hopping between the $p_z$ oribtals of A and B sublattices yields the $\pi$-character Dirac cone on the honeycomb lattice, while the hopping among the three $\sigma^\ast$ oribtals yields the $\sigma$-character Dirac cone on the kagome lattice (Three bond centers of D, E and F are taken as the lattice sites). The yellow shade denotes the unit cell.}
\end{figure}

We further calculate the electronic structure (See Fig. 3(d)) for the configuration as the inset of Fig. 2(e) but with H removed, in order to clarify the roles of honeycomb lattice and hydrogenation. Two Dirac cones appear, both locating at the $K$-point. The LDOS analysis shows that the upper Dirac cone (Green lines) is originated from $p_x$, $p_y$ and $s$ orbitals while the lower one (Red lines) from $p_z$ orbitals. The upper Dirac cone is very similar to those in Figs. 2(e) and 2(f) irrespective of the contributed orbitals or band shape. Whereas the lower Dirac cone is more similar to that in graphene\cite{usprl}. These strongly imply that the planar honeycomb structure separates the in-plane ($p_x$, $p_y$ and $s$) and out-of-plane ($p_z$) orbitals, forming their own Dirac cones with $\sigma$ and $\pi$ characters. Hydrogenation with the delocalized $\pi$ orbitals quenches the $\pi$-character Dirac cone, leaving the $\sigma$ Dirac cone intact. As well, the two more electrons contributed by H raise the Fermi energy to the Dirac point. In a word, among the five valence electrons of phosphorus, three forms the in-plane $\sigma$ bonds and the one from $p_z$ is localized by saturation, leaving the residual one on the $\sigma^\ast$ levels and forming an in-plane nature Dirac cone. These are summarized in Fig. 4(a) as the energetic scheme of the hydrogenated phosphorene.

The next question is: why and how the $\sigma$-character Dirac cone forms, and also at $K$-point. It is well known to us that the $\pi$-character Dirac cone is originated from the hopping between $p_z$ orbitals on a honeycomb lattice (See Fig. 4(b)), which can be obtained by directly solving a 2 $\times$ 2 spinless matrix. Whereas for the $\sigma$-character Dirac cone, it represents the electron hopping among the three $\sigma^\ast$ bonds as illustrated in Fig. 4(a), and the effective Hamiltonian has to be described by a 3 $\times$ 3 spinless matrix. If the P-P bond centers are viewed as lattice sites, it can be seen that a kagome lattice is formed as shown in Fig. 4(b) (The lattice sites are denoted by D, E and F). A $K$-point Dirac cone is then the nature consequence of the kagome lattice\cite{LiuGC}. Note that such a kagome lattice is based on the assumption of bond center as the lattice site and hence it possesses the dynamic feature to some extent, which may be responsible for the anisotropy of the Dirac cone and the absence of flat band. This is different from graphene or silicene, where the $\sigma^\ast$ bonds are fully unoccupied with no electron hopping, hence the absence of $\sigma$-character Dirac cone.

In light of such a physics for the unique $\sigma$ Dirac cone, it should be very robust against saturated material species and site (double- or single-side), provided that the injected electron eliminates the $\pi$ Dirac cone. To confirm this, we perform systematical calculations in double-side fluorinated, single-side hydrogenated and single-side fluorinated monolayer blue phosphorene with different $a$. We find that the aforementioned Dirac physics remains valid in these systems. We show the results of double-side fluorinated ($a$=4.1 \AA) and single-side hydrogenated ($a$=4.7 \AA) cases in Figs. 5(a) and 5(b), respectively. We can see clearly the Dirac cone spectra resembling that in double-side hydrogenated system. Notably, the P-P bond length in $a$=4.1 \AA\ double-side fluorinated system is just 2.37 \AA, corresponding to an elongation of 4.4\% compared with that in pristine blue phosphorene.

\begin{figure*}[tbp]
\includegraphics[width=0.95\textwidth]{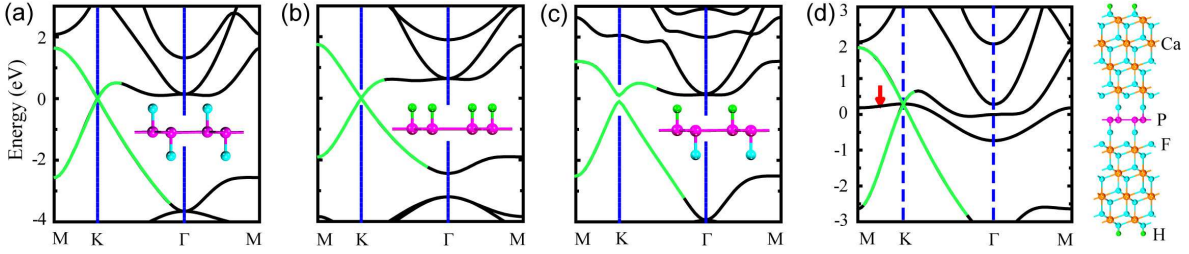}
\caption{\label{Figure 5} Band structures of (a) double-side fluorinated ($a$=4.1 \AA), and (b) single-side hydrogenated ($a$=4.7 \AA) monolayer phosphorene, in both of which there appears the perfect Dirac cone. (c) Gap opening at the Dirac point in one-side hydrogenated and the other side fluorinated monolayer phosphorene ($a$=4.3 \AA). Magenta, cyan and green balls represent the phosphorus, fluorine and hydrogen atoms, respectively. (d) Geometric and electronic structures of double-side saturated phosphorene with F-terminated CaF$_2$, where the Dirac cone remains intact. Note that the two-fold degenerated bands denoted by red arrow are originated from the imperfect passivation of F dangling bonds by H and therefore do not affect the intrinsic transport property. Thick green line is a guide to eye of the Dirac cone.}
\end{figure*}

Application of strain is usually employed to engineer the electronic properties of two-dimensional materials\cite{Rodin,Bissett,Guinea,Pereira,Low,QiJS,ConleyHJ}. Herein, we indeed find the in-plane strain induced gap of 0.35/0.22 eV in single-side hydrogenated/fluorinated phosphorene at $a$=4.4/4.7 \AA. We must emphasize that the gap sizes are not optimized upon $a$ and should set the lower bound in these systems. Even though, it is comparable with the maximum value of 0.25 eV in gate-controlled bilayer graphene\cite{ZhangYB}. So far as is known, it is still a big challenge to open a sizable bandgap in Dirac materials, both graphene\cite{Jang} and topological insulators\cite{usprb}, for their incorporation into the existing semiconductor industry. In addition, we also investigate the system with one-side hydrogenation and the other side fluorination, whose band is shown in Fig. 5(c) ($a$=4.3 \AA). Now an energy gap of 0.19 eV opens at the Dirac point (note: the gap is 0.12 eV when $a$=4.7 \AA). It is worth emphasizing that at $a$=4.3 \AA, the bandgaps are 0.03 and 0 eV, respectively, in double-side hydrogenated and fluorinated systems. Therefore, such a gap opening is related to the AB sublattice symmetry breaking as in graphene. Both in-plane strain and symmetry-breaking can yield a bandgap in our system, making the manipulation more flexible.

Finally, let us discuss the experimental possibility. Since the discovery of graphene, tremendous efforts have been made to search for similar two-dimensional materials. Recently, free-standing single-atomic thick iron membranes suspended in graphene pores have been experimentally demonstrated with a square lattice at room-temperature\cite{Zhao}. Silicene has also been synthesized by depositing silicon on metal surfaces that do not interact strongly with the silicon atoms\cite{Silicene}. In fact, hydrogenation of phosphorene leads to an energy release of $\sim$3 eV per H atom for the systems with Dirac cones, nearly irrespective to $a$. The value is even a little larger than that for hydrogenated graphene\cite{HGra}, implying a good thermodynamic stability of the hydrogenated phosphorene. However, a Jahn-Teller-like structural distortion may anticipate due to the single electron occupation on three degenerate $\sigma^\ast$ oribtals. Our phonon dispersion calculations suggest that either double- or single-side hydrogenation or fluorination might not stabilize the (quasi-)planar honeycomb monolayer phosphorene due to the observation of imaginary frequency. Although the phonon calculations under so large strains are not as reliable as that at equilibrium conditions, it is highly possible that these systems may not exist individually at the finite temperature, particularly as a freestanding form.

The fourfold-coordinated feature, i.e., necessary passivation out-of-plane, provides us with an alternative opportunity because the surface dangling bonds of the semiconducting materials can be utilized to anchor the P atoms and suppress the fluctuation. Three concerns should be considered about the material: (1) saturating the $p_z$ orbital of P; (2) enforcing the P atoms in the honeycomb arrangement; and (3) having an appropriate energy alignment to enable the Dirac cone residing in the gap of the host semiconductor. Given the existence of natural compounds between P and F as well as the huge bandgap ($>$10 eV), CaF$_2$ is taken as an example for illustration and the corresponding geometric and electronic structures are shown in Fig. 5(d). It is noted that the system can be obtained by substituting one layer of Ca atoms with P atoms. Obviously, the typical $\sigma$-character Dirac cone, as in hydrogenated or fluorinated phosphorene, remains unchanged as highlighted by thick green lines. The two-fold degenerate bands (denoted by the red arrow) around the Fermi level arise from the imperfect saturation of F dangling bonds by H on the other side and thus do not reveal the intrinsic physics. In this way, phosphorene can be easily incorporated into the semiconductor industry and importantly, such a sandwich structure also conquers the fragility of phosphorene when exposure to the air\cite{Castellanos-Gomez}. Anyway, stabilizing the $sp^2$ configuration of phosphorus is undoubtedly important and deserves a further investigation.

\section{Conclusion}
In conclusion, we show that Dirac cones can be engineered in phosphorene with fourfold-coordinated phosphorus atom, potentially bringing about many intriguing Dirac-fermions related phenomena. Substantially different from the out-of-plane $\pi$-character in graphene, here obtained Dirac cone exhibits the in-plane $\sigma$-character, with Fermi velocities of 0.91/1.23 times that of graphene along $\Gamma$K/KM direction, which not only enriches the Dirac physics but also offers an excellent opportunity to tune the Dirac point gap via in-plane strain besides the breaking of AB sublattice symmetry. The fourfold-coordinated feature provides us with a new but important freedom for modulating the properties and incorporating into the existing semiconductor industry.

\acknowledgments
We would like to thank H. Q. Huang for the helpful discussion. This work was supported by the National Natural Science Foundation of China (Grant Nos. 11304053).


\begin{references}
\bibitem{Novoselov} Novoselov K S, Fal'ko V I, Colombo L, Gellert P R, Schwab M G and Kim K 2012 \emph{Nature} \textbf{490} 192-200

\bibitem{Avouris} Avouris B, Chen Z H and Perebeinos V 2007 \emph{Nat. Nanotechnol.} \textbf{2} 605-15

\bibitem{Loh} Loh K P, Bao Q L, Eda G and Chhowalla M 2010 \emph{Nat. Chem.} \textbf{2} 1015-24

\bibitem{usprl} Li Y C, Chen P C, Zhou G, Li J, Wu J, Gu B -L, Zhang S B and Duan W H 2012 \emph{Phys. Rev. Lett.} \textbf{109} 206802

\bibitem{Guinea} Guinea F, Katsnelson M I and Geim  A K 2010 \emph{Nat. Phys.} \textbf{6} 30-3

\bibitem{Low} Low T and Guinea F 2010 \emph{Nano Lett.} \textbf{10} 3551-4

\bibitem{Pereira} Pereira V M, Castro Neto A H and Peres N M R 2009 \emph{Phys. Rev. B} \textbf{80} 045401

\bibitem{Reich} Reich E S 2014 \emph{Nature} \textbf{506} 19

\bibitem{LiL} Li L K, Yu Y J, Ye G J, Ge Q Q, Ou X D, Wu H, Feng D L, Chen X H and Zhang Y B 2014 \emph{Nat. Nanotechnol.} \textbf{9} 372-7

\bibitem{LiuH} Liu H, Neal A T, Zhu Z, Luo Z, Xu X F, Tom\'{a}nek D and Ye P D 2014 \emph{ACS Nano} \textbf{8} 4033-41

\bibitem{Xia} Xia F N, Wang H and Jia Y C 2014 \emph{Nat. Commun.} \textbf{5} 4458

\bibitem{Castellanos-Gomez} Castellanos-Gomez A \emph{et al} 2014 \emph{2D Mater.} \textbf{1} 025001

\bibitem{Koenig} Koenig S P, Doganov R A, Schmidt H, Castro Neto A H and \"{O}zyilmaz B 2014 \emph{Appl. Phys. Lett.} \textbf{104} 103106

\bibitem{Qiao} Qiao J, Kong X, Hu Z -X, Yang F and Ji W 2014 \emph{Nat. Commun.} \textbf{5} 4475

\bibitem{blueZhu} Zhu Z and Tom\'{a}nek D 2014 \emph{Phys. Rev. Lett.} \textbf{112} 176802

\bibitem{Fei} Fei R X and Yang L 2014 \emph{Nano Lett.} \textbf{14} 2884

\bibitem{Peng} Peng X H, Wei Q and Copple A 2014 \emph{Phys. Rev. B} \textbf{90} 085402

\bibitem{Zeng} Guo H Y, Lu N, Dai J, Wu X J and Zeng X C 2014 \emph{J. Phys. Chem. C} \textbf{118} 14051

\bibitem{Yao} Lee J, Wang W L and Yao D X 2014 Preprint at http://arxiv.org/abs/1403.7858v1

\bibitem{Rudenko} Rudenko A N and Katsnelson M I 2014 \emph{Phys. Rev. B} \textbf{89} 201408

\bibitem{Silicene} Vogt P, De Padova P, Quaresima C, Avila J, Frantzeskakis E, Asensio M C, Resta A, Ealet B and Le Lay G 2012 \emph{Phys. Rev. Lett.} \textbf{108} 155501

\bibitem{PAW} Bl\"{o}chl P E 1994 \emph{Phys. Rev. B} \textbf{50} 17953-79

\bibitem{PBE} Perdew J P, Burke K and Ernzerhof M 1996 \emph{Phys. Rev. Lett.} \textbf{77} 3865

\bibitem{vasp} Kresse G and Furthm\"{u}ller J 1996 \emph{Phys. Rev. B} \textbf{54} 11169-86

\bibitem{Wallace} Wallace P R 1947 \emph{Phys. Rev.} \textbf{71} 622

\bibitem{Kane} Kane C L and Mele E J 2005 \emph{Phys. Rev. Lett.} \textbf{95} 146802

\bibitem{Rycerz} Rycerz A, Tworzydlo J and Beenakker C W J 2007 \emph{Nat. Phys.} \textbf{3} 172-5

\bibitem{Tahir} Tahir M and Schwingenschl\"{o}gl U 2013 \emph{Sci. Rep.} \textbf{3} 1075

\bibitem{LiuGC} Liu G C, Zhang P, Wang Z G and Li S S 2009 \emph{Phys. Rev. B} \textbf{79} 035323

\bibitem{Rodin} Rodin A S, Carvalho A and Castro Neto A H 2014 \emph{Phys. Rev. Lett.} \textbf{112} 176801

\bibitem{Bissett} Bissett M A, Tsuji M and Ago H 2014 \emph{Phys. Chem. Chem. Phys.} \textbf{16} 11124

\bibitem{QiJS} Qi J S, Qian X F, Qi L, Feng J, Shi D N and Li J 2012 \emph{Nano Lett.} \textbf{12} 1224-8

\bibitem{ConleyHJ} Conley H J, Wang B, Ziegler J I, Haglund R F, Pantelides S T and Bolotin K I 2013 \emph{Nano Lett.} \textbf{13} 3626-30

\bibitem{ZhangYB} Zhang Y B, Tang T T, Girit C, Hao Z, Martin M C, Zettl A, Crommie M F, Shen Y R and Wang F 2009 \emph{Nature} \textbf{459} 820-3

\bibitem{Jang} Jang M S, Kim H, Son Y -W, Atwater H A and Goddard III W A 2013 \emph{Proc. Natl. Acad. Sci.  U.S.A.} \textbf{110} 8786-9
\bibitem{usprb} Li Y C, Tang P Z, Chen P C, Wu J, Gu B -L, Fang Y, Zhang S B and Duan W H 2013 \emph{Phys. Rev. B} \textbf{87} 245127

\bibitem{Zhao} Zhao J, Deng Q M, Bachmatiuk A, Sandeep G, Popov A, Eckert J and R\"{u}mmeli M H 2014 \emph{Science} \textbf{343} 1228-32

\bibitem{HGra} Nechaev Y S and Veziroglu T N 2013 \emph{Open Fuel Cell J.} \textbf{6} 21-39


\end{references}
\end{document}